\documentclass[a4paper]{jpconf}
\usepackage{graphicx}

\def\be {\begin{equation}}
\def\ee {\end{equation}}

\begin{document}
\title{Heavy Flavor Suppression: Boltzmann vs Langevin }

\author{S. K. Das$^{1,2}$, F. Scardina $^{1,2}$, S. Plumari$^{1,2}$ and V. Greco$^{1,2}$}

\address{$^1$ Department of Physics and Astronomy, University of Catania, Via S. Sofia 64, I-
95125 Catania, Italy}

\address{$^2$ Laboratori Nazionali del Sud, INFN-LNS, Via S. Sofia 62, I-95123 Catania, Italy}

\ead{santosh@lns.infn.it}

\begin{abstract}
The propagation of heavy flavor through the quark gluon plasma has been
treated commonly within the framework of Langevin dynamics, i.e. assuming 
the heavy flavor momentum transfer is much smaller than the light one. 
On the other hand a similar suppression factor $R_{AA}$ has been observed 
experimentally for light and heavy flavors. We present a thorough study 
of the approximations involved by Langevin equation by mean of a direct 
comparison with the full collisional integral within the framework of 
Boltzmann transport equation. We have compared the results obtained in 
both approaches which can differ substantially for charm quark leading to 
quite different values extracted for the heavy quark diffusion coefficient. 
In the case of bottom quark the approximation appears to be quite reasonable.
\end{abstract}

\section{Introduction}
The experimental efforts at Relativistic Heavy Ion Collider
(RHIC) and Large Hadron Collider (LHC) energies is to
create and study the properties of matter called quark gluon plasma (QGP). 
The heavy flavors, namely, charm and bottom quarks
are particularly playing a vital role to serve this purpose. 
The most common approach to study heavy flavors propagation in QGP 
is the Fokker-Planck~\cite{hfr,BS,DKS,moore,rappv2,rappprl,gossiauxv2,hiranov2,Das,alberico,jeon,bass,hees} 
equation that can be realized from the Boltzmann
equation which constitutes a significant simplification {\it i.e.} 
the heavy flavor momentum transfer is small or the scattering are 
sufficiently forward peaked. Thus it is a approximation of the full collision 
term of the Boltzmann equation. Such an approximation is expected to be 
asymptotically valid for $m/T \rightarrow \infty$, here we present a first study 
of the validity of such an approximation for b and c quark at temperature 
typically reached in ultrarelativistic heavy ion collisions at LHC.
\section{Relativistic Langevin Equation}
The relativistic Langevin equations of motion corresponds to evolve the particles in the phase space 
according to the following equations~\cite{hfr},
\begin{eqnarray}
 dx_i=\frac{p_i}{E×}dt, \nonumber \\
 dp_i=-\Gamma p_i dt+C_{ij}\rho_j\sqrt{dt}
\end{eqnarray}
where $\Gamma$ and $C_{ij}$ describe the  drag force and the
stochastic force in terms of independent Gaussian-normal 
distributed random variables $\rho$, which obey 
$<\rho_i \rho_j>=\delta_{ij}$ and $<\rho_i>=0$, respectively. 
To study the momentum evolution of charm and bottom quarks, we use the Ito discretization. 
\section{Relativistic Boltzmann Transport Equation}
The relativistic Boltzmann equation can be written as follows~\cite{greco_cascade,Greiner_cascade}:
\begin{equation}
p^{\mu} \partial_{\mu} f(x,p)=
{\cal C}(x,p)
\label{BV-equation}
\end{equation}
where $\mathcal{C}(x,p)$ is the Boltzmann collision integral, the main ingredient of the cascade
codes.

If we define $\omega(p,k)$, the rate of collisions
which change the heavy quark momentum
from $p$ to $p-k$, then we have~\cite{BS}
\be
{\cal C}(x,p) = \int d^3k \left[ \omega(p+k,k)f(p+k) 
- \omega(p,k)f(p) \right]
\label{expeq_00}
\ee
The first term in the integrand represents the gain of probability
through collisions which knock the heavy quark
into the volume element, and the second
term represents the loss out of that volume element.
If we expand $\omega(p+k,k)f(p+k)$ around $k$,
\be
\omega(p+k,k)f(p+k) \approx \omega(p,k)f(p) +k \frac{\partial}{\partial p} (\omega p) 
+\frac{1}{2}k_ik_j \frac{\partial^2}{\partial p_i \partial_j} (\omega p)
\label{expeq_000}
\ee
We obtain the Fokker Planck/Langevin Equation. In the following we will discuss a 
first comparison between the full Boltzmann equation and the Langevin equation 
coming from  Eq.~\ref{expeq_000}.
\section{Numerical results and discussion}
The basic ingredients required for solving the Langevin equation are the drag and diffusion coefficients 
and the initial momentum distributions of heavy quarks. We are using $f_{t=0}^c=4.1*10^2/(0.7+0.09p)^{15.44}$ 
and $f_{t=0}^b=1/(57.74+p^2)^{5.04}$ as the initial momentum distribution of the charm and bottom quarks respectively.
For the sake of comparison, we 
are solving the Langevin equation in a box where the bulk consists of only gluon at T=0.4 GeV. 
The elastic collisions of heavy quarks with gluon has been considered within the pQCD framework to calculate 
the drag and diffusion coefficients. In the present calculation we used $\alpha_s=0.35$ and the 
Debye screening mass as $m_D=gT$. It can be mentioned that the experimental data on the nuclear 
suppression factor and the elliptic flow of nonphotonic electrons provide the best agreement
to the experimental data for the spatial diffusion coefficient $D_x=6/2\pi T$~\cite{bass}. Keeping this in mind 
we enhanced the pQCD cross sections by a factor 2 to match our diffusion coefficients as $D_x=6/2\pi T$, 
so that we could extract conclusion of phenomenological interest from our calculation. 
On the other hand the relativistic Boltzmann equation also solved in a identical environment as the 
Langevin equation with the enhanced pQCD cross sections by a factor 2, in a Montecarlo cascade~\cite{greco_cascade} 
based on the stochastic interpretation of the transition rate discussed in Ref~\cite{Greiner_cascade}.

\begin{figure}[ht]
\begin{center}
\includegraphics[width=17pc,clip=true]{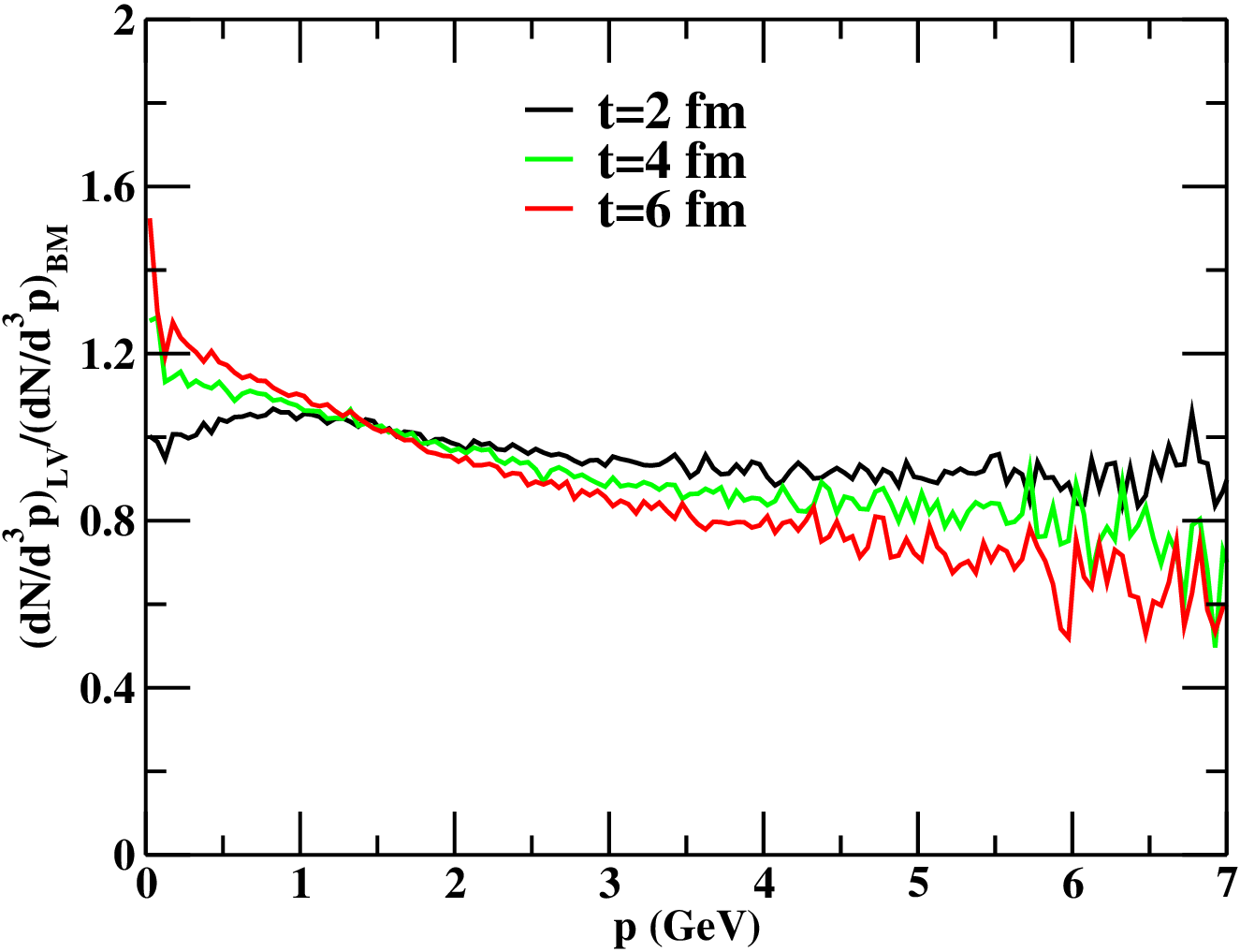}\hspace{2pc}
\caption{\label{label}Ratio between the Langevin (LV) and Boltzmann (BM) spectra for charm quark 
as a function of momentum at different time}
\label{fig1}
\end{center}
\end{figure}

\begin{figure}[ht]
\begin{center}
\includegraphics[width=17pc,clip=true]{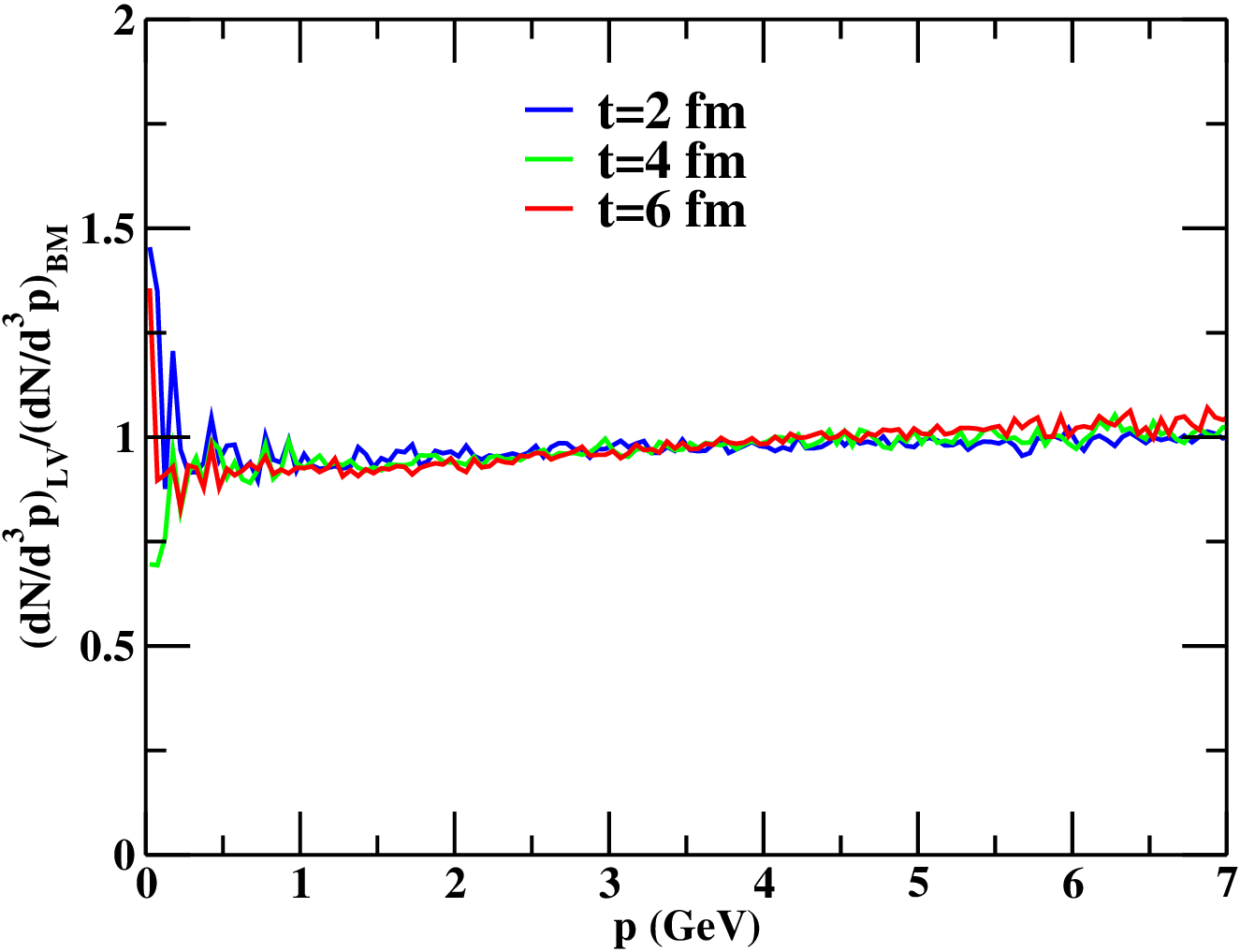}\hspace{2pc}
\caption{\label{label}Ratio between the Langevin (LV) and Boltzmann (BM) spectra for bottom quark
as a function of momentum at different time}
\label{fig2}
\end{center}
\end{figure}

\begin{figure}[ht]
\begin{center}
\includegraphics[width=17pc,clip=true]{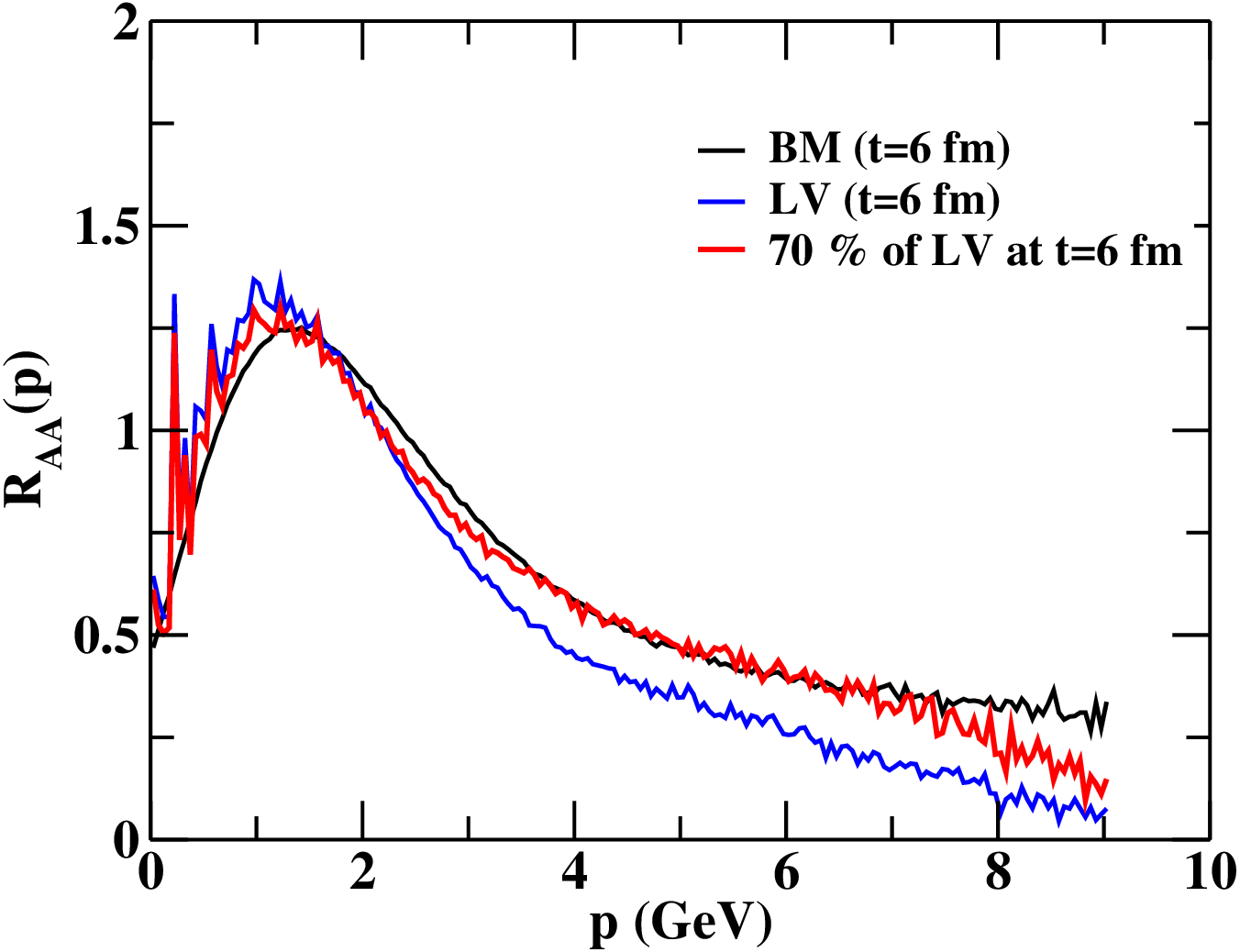}\hspace{2pc}
\caption{\label{label}The nuclear suppression factor, $R_{AA}$ as a function of 
momentum from the Langevin (LV) equation and Boltzmann (BM) equation for charm quark in a box at T=0.4 GeV}
\label{fig3}
\end{center}
\end{figure}

In Fig~\ref{fig1} the ratio of Langevin to Boltzmann spectra for the charm quark has been displayed 
as a function of momentum 
at different times to quantify how much the ratio deviates from 1. We started the simulation 
at $t=0$ with the same initial distribution for both Lanvevin and Boltzmann equations which leads 
to a ratio 1. So any deviation from 1 would reflect how much the Langevin deviations differ from 
the Boltzmann as a approximation. From Fig~\ref{fig1} it is observed that for $t=4$ fm the deviation 
of Langevin from Boltzmann is around $25\%$ and for $t=6$ fm the deviation is around $35\%-40\%$ at $p=7$ GeV,  
which suggests Langevin approach overestimates the interaction considerably due to approximation it involved. 
we have seen that the effect can be significantly larger or smaller depending on the value of the 
drag and diffusion coefficients and/or on the angular depndence of the collisions.
The same quantity is depicted in Fig~\ref{fig2}  as a function momentum at different times for bottom quark. 
From Fig~\ref{fig2} it is shown that the ratio is almost unity at all the times considered in the manuscript. 
Therefore, for bottom quark the Langevin approach is really a good approximation of Boltzmann equation.

We calculate the nuclear suppression factor, $R_{AA}$,  using our initial $t=0$ 
and final $t=t_f$ charm quark distribution as $R_{AA}=\frac{f_{t=f}}{f_{t=0}×}$. 
The nuclear suppression factor, $R_{AA}$, has been displayed in Fig~\ref{fig3} as 
a function of momentum from both Langevin and Boltzmann side at $t=6$ fm. 
From Fig~\ref{fig3} it is observed that the nuclear suppression factor differ substantially 
from Langevin to Boltzmann. Since the diffusion coefficient is very important for the phenomenological 
study, it will be very useful from phenomenological point of view to study how much the diffusion 
coefficient change from Langevin side to reproduce the same nuclear suppression factor of Boltzmann 
equation. It is observed that (Fig~\ref{fig3}) we need to change (reduced) the diffusion coefficient of Langevin 
equation by $30\%$ to get similar nuclear suppression factor as of Boltzmann equation, although it is not anyway possible 
to reproduce the same momentum dependence. Heavy flavor supression within Boltzmann tranport approach can be 
found in Ref.~\cite{gre}(see also Ref.~\cite{you}) .     
\section{SUMMARY AND CONCLUSIONS}
We present a thorough study 
of the approximations involved by Langevin equation by mean of a direct 
comparison with the full collisional integral within the framework of 
Boltzmann transport equation in a box where the bulk consists of only gluon 
at T=0.4 GeV. We found that the Langevin approach is a good approximation 
for bottom quark where as for charm quark Langevin approach deviates 
of about a $30\%$ at intermediate momentum for a time evolution of 5-6 fm/c typical 
of the QGP created in heavy-ion collisions respect to the spectra calculated 
from the solution of the full Boltzmann transport equation. It has also been found that 
to get a similar suppression factor form both the approach we need to reduce 
the diffusion coefficient of the Langevin approach by around $30\%$. 
These results can have significant effects on the heavy ion phenomenology 
at RHIC and LHC energies. Calculation for a realistic fireball evolution 
as created in ultra-relativistic heavy-ion collisions is undergoing.

\section*{Acknowledgements} 
We acknowledge the support by the ERC StG under the QGPDyn
Grant n. 259684

\end{document}